\documentstyle[12pt]{article}
\newcommand{\be}{\begin{equation}}
\newcommand{\ee}{\end{equation}}
\newcommand{\beq}{\begin{eqnarray}}
\newcommand{\eeq}{\end{eqnarray}}
\newcommand{\bea}[2]{\be\label{#2}\begin{array}{#1}}
\newcommand{\eea}{\end{array}\ee}

\def\Cb{{\rm \bf C}}

\def\det{\,{\rm det}\, }

\def\Im{\,{\rm Im}\, }

\def\({\left(}
\def\){\right)}
\def\[{\left[}
\def\]{\right]}
\def\p{\partial}
\def\11{1\!\! 1}

\def\hf{{1\over 2}}

\def\eps{\varepsilon}

   \def\CD {{\cal D}}

   \def\CG {{\cal G}}

   \def\CL {{\cal L}}
   \def\CM {{\cal M}}

   \def\CV {{\cal V}}
   \def\CW {{\cal W}}

\newcommand{\tE}{\lefteqn{\smash{\mathop{\vphantom{<}}\limits^{\;\sim}}}E}
\newcommand{\tP}{\lefteqn{\smash{\mathop{\vphantom{<}}\limits^{\;\sim}}}P}
\newcommand{\tQ}{\lefteqn{\smash{\mathop{\vphantom{<}}\limits^{\;\sim}}}Q}
\newcommand{\Et}{\lefteqn{\smash{\mathop{\vphantom{\Bigl(}}\limits_{\sim}
\atop \ }}E}
\newcommand{\Pt}{\lefteqn{\smash{\mathop{\vphantom{\Bigl(}}\limits_{\sim}
\atop \ }}P}
\newcommand{\Qt}{\lefteqn{\smash{\mathop{\vphantom{\Bigl(}}\limits_{\sim}
\atop \ }}Q}

\newcommand{\tNn}{\lefteqn{\smash{\mathop{\vphantom{\Bigl(}}\limits_{\sim}
\atop \ }}{\cal N}}

\newcommand{\SA}{{\cal A}}

\newcommand{\tPb}{{\tP_{\smash{(\beta)}}}}

\newcommand{\tPp}{{\tP_{\smash{(+)}}}}
\newcommand{\tPm}{{\tP_{\smash{(-)}}}}
\newcommand{\tPpm}{{\tP_{\smash{(\pm)}}}}
\newcommand{\tPmp}{{\tP_{\smash{(\mp)}}}}

\newcommand{\Ptp}{{\Pt_{\smash{(+)}}}}
\newcommand{\Ptm}{{\Pt_{\smash{(-)}}}}
\newcommand{\Ptpm}{{\Pt_{\smash{(\pm)}}}}

\newcommand{\Ap}{{A^{\smash{(+)}}}}
\newcommand{\Am}{{A^{\smash{(-)}}}}
\newcommand{\Apm}{{A^{\smash{(\pm)}}}}
\newcommand{\Amp}{{A^{\smash{(\mp)}}}}

\newcommand{\SAp}{{\SA^{\smash{(+)}}}}
\newcommand{\SAm}{{\SA^{\smash{(-)}}}}
\newcommand{\SApm}{{\SA^{\smash{(\pm)}}}}

\newcommand{\Xp}{\(_{+}{X_{\smash{-}}}\)}
\newcommand{\Xm}{\(_{-}{X_{\smash{+}}}\)}
\newcommand{\Xpm}{\(_{\pm}{X_{\smash{\mp}}}\)}

\newcommand{\nd}{{\cal N}_D}

\newcommand{\Ref}[1]{(\ref{#1})}

\def\R{R}

\def\Rp{{R_{(+)}}}
\def\Rm{{R_{(-)}}}
\def\Rpm{{R_{(\pm)}}}
\def\Rmp{{R_{(\mp)}}}

\newcommand{\tl}{\bf }

\def\plabel#1{\label{#1}}

\begin{document}
%
%

\begin{center}

{\Large \bf Reality conditions for Ashtekar gravity from
Lorentz-covariant formulation }

\

\renewcommand{\thefootnote}{\fnsymbol{footnote}}

Sergei Alexandrov

\setcounter{footnote}{0}
\renewcommand{\thefootnote}{\arabic{footnote}}

\

{\it
Institute for Theoretical Physics \& Spinoza Institute
\\
Utrecht University, Postbus 80.195, 3508 TD Utrecht, The Netherlands
}



\end{center}

\begin{abstract}

We show the equivalence of the Lorentz-covariant canonical
formulation considered for the Immirzi parameter $\beta=i$ to the
selfdual Ashtekar gravity. We also propose to deal with the
reality conditions in terms of Dirac brackets derived from the
covariant formulation and defined on an extended phase space which
involves, besides the selfdual variables, also their anti-selfdual
counterparts.

\end{abstract}


%
\section{Introduction}
%

The complex Ashtekar variables \cite{Ashtekar:1987gu,Ashtekar:1986yd,Ashtekar:1991hf}
were at the origin of the canonical quantization
program of general relativity, which grew up later into the loop approach to quantum
gravity \cite{loops1,loops2,Rovbook}.
However, the modern loop quantization is based on the use of
a real version of the Ashtekar variables \cite{Barbero:1994an,Barbero:1994ap}.
The initial complex variables were given up due to the problem associated
with implementation of some reality conditions needed to ensure that one describes
the real Einstein gravity \cite{Bengtsson:1989pe,Immirzi:1992ar}.
Despite many efforts to understand these
conditions properly, their status in quantum theory remained obscure.
Therefore, when it was realized that the real Barbero variables
preserve the main advantages of the Ashtekar variables and still allow
a quantization {\it \`a la} loops, the interest has moved to this direction.

However, there are at least two big differences between the complex and the real variables.
First, the former describe a theory with the Lorentz gauge group, whereas in the latter
case the gauge group is reduced to SU(2) \cite{Immirzi:1996di,Immirzi:1996dr}.
Second, the Ashtekar connection is
a pull-back of a spacetime connection, whereas the real Barbero connection
is not \cite{Samuel:2000ue}.
These differences warn us that the passage to the
real variables may not be so harmless as it seems from the first sight.

In fact, in some of our previous works \cite{SAcon,AlLiv}
it was argued that the loop quantization
based on the real Barbero variables is very likely anomalous.
The main reason for this is just the second fact mentioned above, from which
it follows that the Barbero connection does not transform as a true connection
under the time diffeomorphisms.
Therefore, one expects an anomaly in the diffeomorphism symmetry
at the quantum level. In particular, it was argued that
one of the manifestations of such an anomaly is the appearance of
the Immirzi parameter in physical results,
such as spectra of geometric quantities \cite{area1,ALarea,Ashtekar:1997fb}.

All these conclusions were obtained in the framework of the
so-called covariant loop quantization originating from a canonical
formulation explicitly covariant under the full Lorentz gauge
group \cite{SA}. Using this formulation, it was shown there is
only one connection, which transforms properly under all classical
symmetries (four diffeomorphisms and six local Lorentz
transformations) and simultaneously diagonalizes the area operator
\cite{SAcon}. As expected, it does not coincide with the Barbero
connection and leads to the results different from the ones found
in loop quantum gravity with SU(2) gauge group. In particular, the
area spectrum does not depend on the Immirzi parameter, is given
by the Casimir operator of SO(3,1) and, therefore, continuous
\cite{AV,SAcon}.

We see that in many respects the Lorentz-covariant canonical formulation
is quite similar to the original complex formulation of Ashtekar.
Indeed, they both preserve the full Lorentz gauge symmetry and are
based on connections which are pull-backs of true spacetime connections.
Moreover, one of the key ingredients of the covariant formulation
is the presence of second class constraints. As it will be shown in this paper,
but also expected on general grounds, they coincide with the reality conditions
of Ashtekar gravity and, as a result, the two formulations are
completely equivalent.

The reality conditions were the main obstacle to quantize Ashtekar gravity.
In the covariant approach the second class constraints, which
are equivalent to the reality conditions, are taken into
account via the Dirac bracket. Although some of the resulting expressions
are quite complicated, in principle, this is enough to implement
the constraints at the quantum level. Thus, one can ask: can one learn
something useful about the reality conditions for Ashtekar gravity
starting from the covariant formulation?

One of the problems to understand the reality conditions was that they cannot be
written as constraints on the phase space variables of Ashtekar gravity
because they involve complex conjugate fields. The complex conjugation brings
out of the phase space since there is no symplectic structure defined on
the conjugate fields. Therefore, although it is clear that the reality conditions
are a kind of second class constraints, it is difficult to make this
statement precise.\footnote{On different points of view on this problem in the literature
see \cite{Bengtsson:1989pe,Morales-Tecotl:1996sh,Pons:1999xt}.
In fact, for the reality conditions in the triad form \cite{Yoneda:1996ph},
which are obtained by fixing the time gauge \cite{Pons:1999xt},
one can define a symplectic structure on the conjugate fields since
they are expressed in terms of the original fields. But this is impossible to do
for the reality conditions in the metric form, which preserve the full Lorentz symmetry.}

The simplest idea to deal with this problem would be to extend
the phase space including the conjugate fields
and defining a symplectic structure on them, which should satisfy some consistency
conditions. The original fields of Ashtekar gravity are the selfdual parts of
the triad and the spacetime connection. The complex conjugate fields are
their anti-selfdual counterparts.
Thus, it is natural to expect that the symplectic structure we are looking for
should be induced from a formulation which involves both selfdual and anti-selfdual
fields. For example, it can be the covariant formulation taken for any value of
the Immirzi parameter $\beta\ne \pm i$. Then Ashtekar gravity should be recovered
in the limit $\beta\to i$.

However, one encounters an immediate problem that, considered for canonical variables,
the latter limit is not well defined: the covariant formulation exists for arbitrary
Immirzi parameter except just these two special values where various expressions become
singular. This can be traced back, of course, to the disappearance of the (anti-)selfdual
variables from the action.

A way to overcome this problem comes from the observation that the algebra of Dirac
brackets written for the spacetime connection diagonalizing the area operator,
which played a crucial role in the covariant quantization, does not depend on the
Immirzi parameter \cite{Alexandrov:2005ar}.
Hence, after we take the second class constraints (the reality
conditions) into account and shift the connection properly, the limit $\beta\to i$
becomes smooth.

However, this is not the end of the story yet. First, one should show
that the selfdual and anti-selfdual parts of the shifted connection
can be associated with the Ashtekar connection and its complex conjugate,
respectively. Second, since the limit involves some not well defined
intermediate steps, it is necessary to check that the resulting Dirac brackets
define a consistent symplectic structure. And finally, one should explain how
this helps to solve the problem of the reality conditions in quantum theory.
This is what we are going to accomplish in this paper.

We start by reviewing some necessary elements of the Lorentz-covariant
canonical formulation. In section \ref{sec_sep} we rewrite it in terms of selfdual
and anti-selfdual variables. Then in section \ref{sec_lim} we take the limit
$\beta\to i$ and obtain the complex Ashtekar gravity with an extended phase space
and with the reality conditions taken into account by means of Dirac brackets.
In section \ref{sec_quant} we comment on the quantization of the resulting theory.
In appendices one can find some details of calculations.

%
\section{Lorentz-covariant canonical formulation}
%
\label{sec_covfor}

The staring point to construct the covariant canonical formulation is
the generalized Hilbert--Palatini action \cite{Holst:1995pc}, which allows to include
arbitrary Immirzi parameter $\beta$:
\be
S_{(\beta)}=\hf \int \eps_{\alpha\beta\gamma\delta}
e^\alpha \wedge e^\beta \wedge (\Omega^{\gamma\delta}+
\frac{1}{\beta}\star\Omega^{\gamma\delta}),
\label{Sd1}
\ee
Here $\Omega^{\gamma\delta}=d\omega^{\gamma\delta}+
{\omega^\gamma}_\alpha \wedge \omega^{\alpha\delta}$ is the curvature of
the spin-connection $\omega^{\alpha\beta}$ and $\star$ is the Hodge operator
acting on the tangent indices $\alpha,\beta,\dots$. The notations we use for other
indices are the following. The indices $i,j,\dots$ from the middle of
the alphabet label the space coordinates, $a,b,\dots$ from the beginning are
$so(3)$ indices in the tangent space and the capitalized letters $X,Y,\dots$
take 6 values and are used to label the components of the adjoint representation
of $sl(2,\Cb)$.

The canonical formulation arises after the $3+1$ decomposition
\be
e^0=Ndt+\chi_a E_i^a dx^i, \quad e^a=E^a_idx^i+E^a_iN^idt.
\plabel{decomp}
\ee
The field $\chi$ appearing in \Ref{decomp}
describes the deviation from the time gauge
$\chi=0$, which is used to obtain the real Barbero formulation.
To write the decomposed action, it is convenient to introduce
the fields $A_i^X$ and $\tPb^i_X$, which belong the adjoint representation
of the Lorentz group and are defined as follows \cite{SA}
\beq
&& A_i^{ X}=(\omega_i^{0a},\frac12 {\eps^a}_{bc}\omega_i^{bc}),
\qquad\qquad
\tPb_X^i=\tP_X^i-\frac{1}{\beta}\tQ_X^i,
\plabel{multHP}\\
&&\quad {\rm where}\qquad  \tP_X^{ i}=(\tE^i_a,{\eps_a}^{bc}\tE^i_b\chi_c),
\qquad
\tQ_X^{ i}=(-{\eps_a}^{bc}\tE^i_b\chi_c,\tE^i_a)
\nonumber
\eeq
and $\tE^i_a=h^{1/2}E^i_a$ $(\sqrt{h}=\det E^a_i)$ is the inverse densitized triad.
The first field is just the space components of the spin-connection
$\omega^{\alpha\beta}$, the field $\tP^i_X$ can be obtained from the
bivector $e^{\alpha}\wedge e^{\beta}$, and $\tQ^i_X$ comes from its
Hodge dual. This fact is encoded in the relation
\be
\tP^i_X=\Pi_X^Y\tQ^i_Y,
\plabel{matR}
\ee
where the matrix $\Pi$ can be considered as a representation of the $\star$ operator
and is defined in appendix \ref{ap_def}. There one can find also the definition
of the Killing form $g^{XY}$ of the $sl(2,\Cb)$ algebra, its structure constants
$f_{XY}^Z$ and various properties satisfied by these matrices and fields.

In terms of the introduced fields and after some redefinition
of the lapse and shift, the decomposed action takes the following form \cite{SA}
\beq
S_{(\beta)} &=&\int dt\, d^3 x (\tPb^i_X\partial A^X_i
+A_{0}^X \CG_X+\nd^i H_i+\tNn H),  \label{dact} \\
\CG_X&=&\partial_i \tPb^i_X +f_{XY}^Z A^Y_i \tPb^i_Z,
\nonumber \label{Gauss} \\
H_i&=&-\tPb^j_X F_{ij}^X,
\nonumber \label{hdiff} \\
H&=&-\frac{1}{2\left(1+\frac{1}{\beta^2}\right)}
\tPb^i_X \tPb^j_Y f^{XY}_Z R^Z_W F_{ij}^W,
\nonumber \label{ham} \\
{\rm where}\ \ F^X_{ij}&=&\partial_i A_j^X-
\partial_j A_i^X+f_{YZ}^X A^Y_i A^Z_j,
\qquad
R^{XY} =g^{XY}-\frac{1}{\beta}\Pi^{XY}.
\nonumber \label{FF}
\eeq
It is clear that  $A_i^X$ and $\tPb^i_X$ form the canonical pair and
$\CG_X,\ H_i$ and $H$ are first class constraints generating the symmetry
transformations. However, there are additional constraints coming from the
fact that not all components of $\tPb^i_X$ are independent. It is easier
to write them in terms of $\tQ^i_X$:
\be
\phi^{ij}=\Pi^{XY}\tQ^i_X\tQ^j_Y=0.
\plabel{phi}
\ee
This constraint is very well known in the BF formulations of gravity
and spin foam models by the name ``simplicity constraint" \cite{BC,Oriti,perez}.
Requiring that $\phi^{ij}=0$ preserved by evolution,
one obtains an additional constraint
\be
\psi^{ij}=2f^{XYZ}\tQ_X^{l}\tQ_Y^{\{ j}\partial_l \tQ_Z^{i\} }
-2(\tQ\tQ)^{ ij }\tQ_Z^{l}A_l^Z+
2(\tQ\tQ)^{l\{i  }\tQ_Z^{j\}}A_l^Z=0.
\plabel{psi}
\ee
Here $(\tQ\tQ)^{ij}=g^{XY}\tQ_X^i\tQ_Y^j$ and symmetrization
$\{\cdot\ \cdot\}$ is taken with the weight $1/2$.
Together $\phi^{ij}$ and $\psi^{ij}$ form a set of second class constraints
and require a modification of the symplectic structure to that of
the Dirac brackets \cite{Dirac}.
As a result, the canonical variables acquire the following
non-trivial commutation relations
\beq
\{ \tPb_X^i,\tPb_Y^j\}_D&=&0,
\nonumber \\
\{ A^X_i,\tPb_Y^j\}_D&=&\delta_i^j\delta^X_Y-\frac12 R^{XZ}
\left(\tQ^j_Z\Qt_i^W+\delta^j_i I_{(Q)Z}^W
\right)g_{WY}, \label{comm} \\
\{ A^X_i,A^Y_j\}_D&=& {\rm complicated}.  \nonumber
\eeq
To write the result, we introduced the so-called inverse fields $\Pt_i^X$ and
$\Qt_i^X$ and the projectors
\be
I_{(P)X}^Y = \tP^{ i}_X\Pt_i^{Y},
\qquad
I_{(Q)X}^Y = \tQ^{ i}_X\Qt_i^{Y}.
\plabel{proj}
\ee
We refer to appendix \ref{ap_def} for their definitions in terms of the triad
$\tE^i_a$ and the field $\chi^a$ as well as for their properties.
The commutator of two connections was not specified since
it will not be necessary here.

The connection $A_i^X$ is not well suited for the loop quantization.
The reason is that its commutator with the triad multiplet $\tP^i_X$ given in
\Ref{comm} is not proportional to $\delta_i^j$ and therefore the area operator
is not diagonal on holonomies of this connection \cite{AV}. However, as it was mentioned
in the introduction, there is a unique spacetime connection which does this job.
It can be obtained from $A_i^X$ by shifting it by a term proportional to the
Gauss constraint:
\be
\SA_i^X=A_i^X + \frac{1}{2\left(1+\frac{1}{\beta^2}\right)}
\R^{X}_{S}I_{(Q)}^{ST}\R_T^Z f^Y_{ZW}\Pt_i^W \CG_Y
= I_{(P)Y}^X \(\delta^{Y}_Z+\frac{1}{\beta}\Pi^{Y}_Z\) A_i^Z
+ R^X_Y \Gamma_i^Y,
\plabel{spconnew}
\ee
where
\be
\Gamma_i^X = \frac12 f^{W}_{YZ}I_{(Q)}^{XY} \Qt_i^Z \p_l \tQ^l_W
+\frac12 f^{ZW}_Y\left((\Qt\Qt)_{ij} I_{(Q)}^{XY}
+\Qt_j^X\Qt_i^Y -\Qt_i^X\Qt_j^Y \right) \tQ^l_Z \p_l \tQ^j_W
\plabel{gam}
\ee
and we used \Ref{psi} to obtain the second equality.
The quantity $\Gamma_i^X$ is nothing else but the SL(2,\Cb) connection compatible
with the metric induced on the 3-dimensional hypersurface \cite{Alexandrov:2005ar}.
At $\chi=0$ it reduces to the connection $\Gamma_i^a(\tE)$
appearing in the definition of the Barbero connection.

In terms of the new connection the Dirac brackets take a simpler form
and do not depend at all on the Immirzi parameter:
\beq
\{\SA_i^X(x),\tP_Y^j(y)\}_D&=&\delta_i^j I_{(P)Y}^X\delta(x,y),
\plabel{commsa} \\
\{ \SA^X_i(x), \SA^Y_j(y)\}_D &=&
\hf \( \Pi^X_{X'} \CM_{ij}^{X'Y}- \CM_{ij}^{XY'}\Pi_{Y'}^Y\)  \delta(x,y).
\plabel{commAAA}
\eeq
Here $\CM_{ij}^{XY}$ is a linear differential operator whose exact expression
can be found in appendix \ref{ap_def}.
An important consequence of \Ref{commsa} is that the field $\chi$ and, therefore,
also the projectors $I_{(P)}$ and $I_{(Q)}$ commute with both $\tP$ and $\SA$.

%
\section{Separation of chiral variables}
%
\label{sec_sep}

This section is purely technical.
Our aim here is to split all variables into the selfdual and anti-selfdual parts.
For this let us introduce the corresponding projectors
\be
\Rpm^{XY}
=\hf\(g^{XY}\mp i\Pi^{XY}\)=\left(
\begin{array}{cc}
1& \mp i \\
\mp i & -1
\end{array}
\right)\frac{\delta^{ab}}{2} ,
\plabel{matirch}
\ee
which satisfy the following properties
\be
\Rp \cdot\Rm=0,
\qquad (\Rpm)^2=\Rpm, \qquad \Pi\cdot \Rpm=\pm i \Rpm.
\plabel{relmatich}
\ee
Applying these projectors to the canonical fields, one obtains
\beq
\Rpm_X^Y \tP^i_Y=\hf(\tPpm^i_a,\pm i \tPpm^i_a), \qquad
\tPpm^i_a=\tE^i_a\mp i{\eps_a}^{bc}\tE^i_b\chi_c, \\
\Rpm^X_Y A_i^Y=\hf(\Apm_i^a, \mp i\Apm_i^a), \qquad
\Apm_i^a=\omega_i^{0a}\pm \frac{i}{2}{\eps^a}_{bc}\omega_i^{bc},
\label{Abich}
\eeq
and $\Rpm \cdot \tQ=\mp i \Rpm\cdot \tP$. Thus, each of the projected fields have only half
of independent components, so one can take
$(\tPp,\Ap)$ and $(\tPm,\Am)$ to be the basic variables.
It is useful to notice also the following relations
\beq
& \R \cdot \Rpm = \(1\mp\frac{i}{\beta}\)\Rpm, &
\nonumber
\\
&
\Rpm\cdot I_{(Q)}\cdot \Rpm=\hf\,\Rpm, \qquad
\Rpm\cdot I_{(P)}\cdot \Rpm=\hf\,\Rpm, &
\plabel{matproj}
\\
& \(\Rpm\cdot I_{(P)}\cdot \Rmp\)^{XY}=\left(
\begin{array}{cc}
1&  \pm i \\
\mp i & 1
\end{array}
\right)\frac{\displaystyle \Xpm^{ab}}{\displaystyle 4}, &
\nonumber
\eeq
where we introduced
\be
\Xpm^{ab}=\frac{\delta^{ab}(1+\chi^2)-2\chi^a\chi^b\mp 2i\eps^{abc}\chi_c}{1-\chi^2}.
\plabel{matX}
\ee
It is easy to check that the matrices $\Xp^{ab}$ and $\Xm^{ab}$ are mutually inverse.

After the splitting of the variables into selfdual and anti-selfdual parts,
the action \Ref{dact} can be written as a sum of two actions. One of them
depends only on the selfdual variables and the other one is a similar action
for the anti-selfdual fields:
\be
S_{(\beta)}=\textstyle\frac{1+i/\beta}{2}\, S_{(+)}
+\textstyle\frac{1-i/\beta}{2}\, S_{(-)},
\plabel{chact}
\ee
where
\beq
S_{(\pm)} &=&\int dt\, d^3 x (\tPpm^i_a\partial \Apm^a_i
+\Apm_{0}^a \CG^{(\pm)}_a+\nd^i H^{(\pm)}_i+\tNn H^{(\pm)}),  \label{dactpm} \\
\CG^{(\pm)}_a&=&\partial_i \tPpm^i_a \pm i{\eps_{ab}}^c\, \Apm^b_i \tPpm^i_c,
\nonumber \label{Gausspm } \\
H^{(\pm)}_i&=&-\tPpm^j_a F^{(\pm)a}_{ij},
\nonumber \label{hdiffpm} \\
H^{(\pm)}&=&\mp\frac{i}{2}
\tPpm^i_a \tPpm^j_b \, {\eps^{ab}}_c \, F^{(\pm)c}_{ij},
\nonumber \label{hampm} \\
F^{(\pm)a}_{ij}&=&\partial_i \Apm_j^a-
\partial_j \Apm_i^a\pm i{\eps^a}_{bc}\, \Apm^b_i \Apm^c_j.
\nonumber \label{FFpm}
\eeq
Thus, the two chiral sectors do not interact with each other and the Immirzi parameter
measures the ``weight" of each sector.
The only non-vanishing Poisson brackets of the chiral variables are
\be
\{\Apm_i^a,\tPpm^j_b \}= \frac{2\delta_i^j\delta^a_b}{1\pm \frac{i}{\beta}}.
\label{Pbchir}
\ee

The two sectors become mixed when one takes into account the second class constraints
\Ref{phi} and \Ref{psi}. Let us also rewrite them in terms of the chiral variables.
For the first constraint one has
\be
2i\phi^{ij}=\tPp^i_a\tPp^j_a-\tPm^i_a\tPm^j_a=0.
\label{phich}
\ee
Since $\tPm$ is the complex conjugate of $\tPp$,
the meaning of this constraint is just that the spatial metric
defined by the selfdual triad, $g_{(+)}^{ij}=\tPp^i_a\tPp^j_a$, is real.
Thus, the first of the second class constraints is nothing else but the first reality
condition in the metric form.

The second constraint can be written (with the use of \Ref{phich}) as
\be
-i\psi^{ij}=\(
i\eps^{abc}\tPp_a^{l}\tPp_b^{\{ j}\partial_l \tPp_c^{i\} }
-g_{(+)}^{ ij }\tPp_a^{l}\Ap_l^a+
g_{(+)}^{l\{i  }\tPp_a^{j\}}\Ap_l^a\) - \Bigl({(+)\to (-) \atop i \to -i }\Bigr)=0.
\label{psich}
\ee
Since this constraint was obtained by commuting $\phi^{ij}$ with the Hamiltonian,
it coincides with the second reality condition which requires the reality of the
spatial metric to be preserved under the time evolution.
Thus, as it was expected, the reality
conditions are identical to the second class constraints of the covariant formulation.

In this formulation the constraints were taken into account by means of Dirac bracket.
Making projection to the two chiral sectors in the commutation relation \Ref{comm},
one finds the following results
for the Dirac brackets of the chiral variables
\beq
\{ \Apm^a_i,\tPpm_b^j\}_D&=&\frac{2\delta_i^j\delta^a_b}{1\pm \frac{i}{\beta}}
-\frac12 \frac{1\mp\frac{i}{\beta}}{1\pm\frac{i}{\beta}}
\left(\tPpm^j_a\Ptpm_i^b+ \delta^j_i \delta^{a}_b\right),
\nonumber \\
\{ \Ap^a_i,\tPm_b^j\}_D&=&
\hf \left(\tPp^j_a\Ptm_i^b+ \delta^j_i {\Xp^a}_b\right),
\label{commchir} \\
\{ \Am^a_i,\tPp_b^j\}_D&=&
\hf \left(\tPm^j_a\Ptp_i^b+\delta^j_i {\Xm^a}_b\right),
\nonumber
\eeq
where we had to introduce
\be
\Ptpm_i^a=\frac{(\delta^a_b-\chi^a\chi_b)\Et_i^b\mp i{\eps^a}_{bc}\tE^b_i\chi^c}{1-\chi^2}
\label{invPp}
\ee
such that $\Ptpm_i^a\tPpm^i_b=\delta^a_b$ and $\Ptpm_i^a\tPmp^i_b={\Xpm^a}_b$.

Finally, we should find the chiral components of the shifted connection \Ref{spconnew}.
A simple calculation gives
\be
\SApm_i^a = \hf\(1\pm\frac{i}{\beta}\)\Apm_i^a+
\(1\mp\frac{i}{\beta}\)\(\hf {\Xpm^a}_b \, \Amp_i^b
+ {\Gamma^{\smash{(\pm)}}}_i^a\),
\plabel{connewch}
\ee
where ${\Gamma^{\smash{(\pm)}}}_i^a $ are the chiral components of \Ref{gam}.
Notice that they cannot be written entirely in terms of the fields of one chirality.
Instead, one has the following property
\be
{\Xp^a}_b{\Gamma^{\smash{(-)}}}_i^b=-{\Gamma^{\smash{(+)}}}_i^a.
\plabel{propchg}
\ee
For the variables \Ref{connewch} the Dirac brackets become
\beq
& \{ \SApm^a_i,\tPpm_b^j\}_D= \delta_i^j \delta^a_b, &
\nonumber \\
& \{ \SAp^a_i,\tPm_b^j\}_D= \delta_i^j {\Xp^a}_b ,
\quad \ \quad
\{ \SAm^a_i,\tPp_b^j\}_D= \delta_i^j {\Xm^a}_b , &
\plabel{comsh}
\\
& \{\SApm^a_i, \SApm^b_j\}_D=0, &
\nonumber \\
& \{\SAp^a_i, \SAm^b_j\}_D= i \(\Rp\cdot \CM_{ij}\cdot \Rm\)^{ab} . &
\nonumber
\eeq

This finishes the preparation for taking the limit corresponding to the complex Ashtekar
gravity, which will be investigated in the next section.

%
\section{Ashtekar gravity with extended phase space}
%
\label{sec_lim}

The complex Ashtekar gravity corresponds to the special case where
the Immirzi parameter is chosen to be $\beta=i$. Setting
this value of $\beta$ in \Ref{chact}, one finds that
\be
S_{(i)}=S_{(+)}.
\ee
Of course, $S_{(+)}$ coincides with the usual Ashtekar action
\cite{Ashtekar:1987gu,Ashtekar:1986yd,Ashtekar:1991hf}.
As a result, only the selfdual variables contribute to the action and
all anti-selfdual variables disappear. An immediate consequence of this
is that the Poisson brackets \Ref{Pbchir} of the anti-selfdual variables become divergent.

The situation does not become better when one takes into account the second class constraints
relating the selfdual and anti-selfdual fields and considers the corresponding Dirac bracket.
Indeed, the results \Ref{commchir} show that the Dirac bracket of the anti-selfdual parts of
the canonical connection and the triad still diverges at $\beta=i$.
Besides, we did not consider the commutator of two connections which also can contain
some divergences.
Thus, there is no consistent symplectic structure which can be defined on the space of
$(\tPp,\Ap)$ and $(\tPm,\Am)$.

Nevertheless, let us consider instead the phase space spanned by $(\tPp,\SAp)$ and $(\tPm,\SAm)$.
Remarkably, the Dirac brackets of these variables, given in \Ref{comsh},
do not depend on $\beta$ and therefore are well defined even at the point
corresponding to the complex Ashtekar gravity.
Thus, they represent a good candidate for the symplectic structure we are looking for,
which will allow to implement the reality conditions at the quantum level.

But how could it happen that all divergences disappeared? It is clear that this cannot
be achieved by a simple invertible change of variables. To clarify the situation,
let us consider the expressions for the chiral components of the shifted connection
\Ref{connewch} in terms of the original variables at $\beta=i$. One finds
\beq
\SAp_i^X &= &\Ap_i^a,
\label{poscon} \\
\SAm_i^a &= &
{\Xm^a}_b\, \Ap_i^b + 2{\Gamma^{\smash{(-)}}}_i^a.
\label{negcon}
\eeq
We observe that none of the chiral components depend on $\Am$. This means that the shifted
connection $\SA$ contains 9 components less than the original connection $A$.
The missing components, which are precisely $\Am$ at $\beta=i$, were removed by
means of the second class constraint $\psi^{ij}$
and some part of the Gauss constraint (at $\beta=i$ this is $\CG^{(-)}_a$).
Thus, working at the surface of these constraints,
we simply exclude the corresponding variables from the phase space.

However, now there is another problem. On one hand, $\SAm$ can be expressed through
other variables by means of \Ref{negcon} as\footnote{Note that for $\chi=0$
the relation \Ref{negcons} reduces to the well known second reality condition
in the triad form: $\Im A_i^a=\Gamma_i^a$.
Our approach provides its generalization to the case of the full Lorentz gauge
group.}
\be
\SAm_i^a ={\Xm^a}_b\, \SAp_i^b + 2{\Gamma^{\smash{(-)}}}_i^a .
\label{negcons}
\ee
On the other hand, its Dirac brackets are already defined
in \Ref{comsh}. Thus, there is a non-trivial consistency condition which requires that,
using the expression \Ref{negcons} for $\SAm$ to calculate the Dirac brackets, one obtains
the same results as in \Ref{comsh}. Equivalently, this means that the relation
\Ref{negcons} can be considered as a strong equality or a second class constraint
on the phase space of $(\tPp,\tPm,\SAp,\SAm)$ endowed with the symplectic structure
\Ref{comsh}. We check that this is indeed true in appendix \ref{ap_con}.

Notice that this consistency condition is not ensured by the construction for generic $\beta$.
The problem is that the shifted connection contains terms proportional to $\(1-i/\beta\)\Am$.
At $\beta=i$ such terms do not contribute to the expression for the connection,
but they do contribute to the Dirac brackets since the vanishing factor
$\(1-i/\beta\)$ can be cancelled by the same factor
from the denominator in \Ref{commchir}. The simplest example of such situation is
the Dirac bracket of $\SAp$ with $\tPm$:
although $\SAp$ coincides with $\Ap$ according to \Ref{poscon},
the Dirac brackets \Ref{commchir} and \Ref{comsh} are different.
The remarkable fact is that despite all these problems, which seem
to appear at intermediate steps, the final result, namely the Dirac brackets
\Ref{comsh} together with the constraint \Ref{negcons}, is consistent.

As a result, one gets the following picture. The phase space of Ashtekar gravity
can be extended to include, besides the original selfdual variables $\SAp$ and $\tPp$,
also $\tPm$ and $\SAm$ with the constraint \Ref{negcons} imposed on it (so, in fact,
$\SAm$ can be excluded from the phase space).
The symplectic structure on this extended phase space is defined by the Dirac
brackets \Ref{comsh}. Finally, on the extended phase space one can define the operation
of complex conjugation, which acts according to
\be
\(\tPp\)^*=\tPm, \qquad \(\SAp\)^*=\SAm.
\plabel{comcon}
\ee
It is easy to see that the two structures, the symplectic structure and the complex
conjugation \Ref{comcon}, are mutually consistent, which means that
\be
\{F^*,G^*\}_D=\{F,G\}_D^*.
\plabel{strconst}
\ee

Before using this construction for quantization of general relativity,
one should check two additional conditions. First, it should
ensure the reality of the metric. Second, the complex conjugation
in \Ref{comcon} should agree with the usual one, which acts in the
evident way being written in the original variables (we denote it
by bar):
\be
\overline{\vphantom{\tilde P}\tPp}=\tPm, \qquad
\overline{\vphantom{\tilde P}\Ap}=\Am.
\plabel{conjst}
\ee
The first requirement is fulfilled due to the fact that the symplectic
structure is induced by the Dirac brackets, which take into
account the second class constraints. As we saw above, these
constraints are nothing else but the reality conditions for the
metric. The second condition becomes satisfied if one allows also
to use the Gauss constraint, since in that case one has (see
\Ref{spconnew})
\be
\overline{\vphantom{\tilde P}\SAp}=
\Am \mathop{\approx}\limits^{\phi=\psi=0\vphantom{A\atop B}}_{\CG=0}
\SAm=\(\SAp\)^*.
\plabel{surfcon}
\ee

Thus, using the Gauss constraint and shifting the canonical connection by a term
proportional to it, allows to achieve two things:
the Dirac brackets become well defined and it becomes possible to endow
the extended phase space with a complex conjugation consistent with the usual one.
The resulting structure will be the starting point to discuss the quantization
of Ashtekar gravity in the next section.

%
\section{Quantization}
%
\label{sec_quant}

Quantizing gravity in the loop approach, one chooses the space of connections
as configuration space and the wave functions to be the loop, or the so called
spin network functionals of the connection \cite{loops2,Baez:1995md,Rovelli:1995ac}.
In our case it is natural to take
them to be the functionals of the selfdual connection $\SAp$. Then the variables
$\tPp$ are going to be the operators which are
the usual functional derivatives with respect to $\SAp$. In this way one obtains
the standard loop quantization of the selfdual sector \cite{Ashtekar:1991hf}.

The main problem, which stayed for long time, was how to implement
the reality conditions of Ashtekar gravity in this framework. The idea was
that it can be done by a clever choice of the scalar product on the space
of loop functionals. But no such scalar product have been found.

The picture presented in the previous section suggests a new look at the problem
of the reality conditions. These conditions will be automatically satisfied
as soon as we find an appropriate representation of the algebra of the Dirac brackets
\Ref{comsh} such that the fields, which are complex conjugate according to the rule
\Ref{comcon}, become hermitian conjugate operators.
In other words, the anti-selfdual fields should be realized as operators
hermitian conjugate to the selfdual ones:\footnote{It might be that the connection
itself is not a well defined operator on the Hilbert space as it happens in the
standard loop approach. In this case, the requirement \Ref{comconop} should
be understood for appropriate functions of the connection.}
\be
\(\widehat{\vphantom{\tilde P} \tPp}\)^{\dagger}=\widehat{\vphantom{\tilde P} \tPm},
\qquad
\(\widehat{\vphantom{\tilde A} \SAp}\)^{\dagger}=\widehat{\vphantom{\tilde A} \SAm}.
\plabel{comconop}
\ee
For example, it is trivial to check that in this case the operator of the spatial metric
and, consequently, the area operator would be hermitian operators.
Thus, it is not necessary to deal explicitly with the constraint expressing
the complex conjugate connection in terms of the original variables.
Rather, the problem is moving in the direction of the representation theory
of some complicated algebra.

In fact, the problem of finding the appropriate representation is
still quite non-trivial, especially taking
into account the very non-trivial form of the commutation relation between
the selfdual and anti-selfdual connections \Ref{comsh}.
Indeed, the simplest solution to \Ref{comconop} would be that in the connection
representation, which is extensively used in the loop approach, the two chiral
connections are realized as multiplication operators by complex conjugate
variables. But this contradicts to their non-vanishing commutator.
Nevertheless, the form of the Dirac algebra suggests that may be it is possible
to realize the selfdual and anti-selfdual connections as such multiplication
operators when they act on the functions of only $\SAp$ or $\SAm$, respectively.

Although we do not know a representation of the algebra \Ref{comsh}
where the selfdual connection is chosen as configuration variable, it is easy
to construct a representation with $\tPp$ and $\tPm$ being configuration
variables. In appendix \ref{ap_repr} we show that the following operators
\beq
\widehat{\vphantom{\tilde P} \tPp^i_a}=\tP^i_a, &
\qquad &
\widehat{\vphantom{\tilde A} \SAp_i^a}=
i\frac{\delta}{\delta\tP^i_a}
+i{\Xp^a}_b\,\frac{\delta}{\delta\overline{\vphantom{\tilde P}\tP^i_b}}
+{\Gamma^{\smash{(+)}}}_i^a(\tP,\overline{\vphantom{\tilde P}\tP}),
\label{prep_self} \\
\widehat{\vphantom{\tilde P} \tPm^i_a}=\overline{\vphantom{\tilde P}\tP^i_a}, &
\qquad &
\widehat{\vphantom{\tilde A} \SAm_i^a}=
i\frac{\delta}{\delta\overline{\vphantom{\tilde P}\tP^i_a}}
+i{\Xm^a}_b\,\frac{\delta}{\delta\tP^i_b}
+{\Gamma^{\smash{(-)}}}_i^a(\tP,\overline{\vphantom{\tilde P}\tP}),
\label{prep_anti}
\eeq
which act on the space of functions of $\tP$ and
$\overline{\vphantom{\tilde P}\tP}$ endowed with
the usual scalar product, form the algebra isomorphic to \Ref{comsh}
and satisfy \Ref{comconop}.
This shows that the search for representations of \Ref{comsh} is not hopeless.
Also it may indicate that the so called triad representation, rather
than the connection representation, might be more natural in quantum gravity.

In fact, a similar problem exists in the covariant approach to
the loop quantization where the non-commutativity of the connection (see \Ref{commAAA})
prevents from choosing it as configuration variable. This problem was
either ignored or some tricks were made to achieve the commutativity for
its holonomies \cite{AlLiv}.
In this respect the situation in Ashtekar gravity is more promising.
It allows to consider holonomies of the selfdual or anti-selfdual connection only
and these chiral quantities are commutative.
Therefore, the only problem arises when one considers their mutual commutators.

We conclude that the results of this paper show the similarity of the approaches
based on the Lorentz-covariant formulation and on the complex Ashtekar formulation,
both in the resulting structures as well as in the arising problems. We hope
that they can help each other to solve these problems
and to find the correct way to quantize gravity.

\section*{Acknowledgments}

The author is grateful to Renate Loll and Hanno Sahlmann for interesting discussions.

\appendix

%
\section{Definitions and properties}
%
\label{ap_def}

{\tl Structure constants of the Lorentz algebra:}
\be
\begin{array}{ccc}
f_{A_1 A_2}^{A_3}=0,&
f_{A_1 B_2}^{A_3}=-\eps^{A_1 B_2 A_3},&
f_{B_1 B_2}^{A_3}=0, \\
f_{B_1 B_2}^{B_3}=-\eps^{B_1 B_2 B_3},&
f_{A_1 B_2}^{B_3}=0,&
f_{A_1 A_2}^{B_3}=\eps^{A_1 A_2 B_3}.
\end{array}      \label{algHP}
\ee
Here we split the 6-dimensional index $X$ into a pair of 3-dimensional
indices, $X=(A,B)$, so that $A,B=1,2,3$. The indices $A$ correspond
to the Lorentz boosts, whereas the indices $B$ label the SO(3) subgroup.

{\tl Killing form:}
\be
g_{XY}=\frac14 f_{XZ_1}^{Z_2}f_{YZ_2}^{Z_1},\quad
g^{XY}=(g^{-1})^{XY}, \quad
g_{XY} =\left(
\begin{array}{cc}
\delta_{ab}&0 \\ 0&-\delta_{ab}
\end{array}
\right).
\ee

{\tl Matrix algebra:}
\be
\Pi^{XY}=(\Pi^{-1})^{XY}=\left(
\begin{array}{cc}
0&1 \\ 1&0
\end{array}
\right)\delta_a^b
\qquad
\R^{XY} =\left(
\begin{array}{cc}
1& -\frac{1}{\beta} \\
 -\frac{1}{\beta} & -1
\end{array}
\right)\delta_a^b,
\label{mat}
\ee
The matrices $\Pi^X_Y$, $\R^{X}_{Y}$ and their inverse commute
with each other. Furthermore, they commute with the structure constants
in the following sense:
\be
f^{XYZ'}\Pi_{Z'}^Z=f^{XY'Z}\Pi_{Y'}^Y. \label{com-fp}
\ee
The contraction of two structure constants can be decomposed as follows:
\be
f^T_{XY} f^W_{TZ}=-g_{XZ}\delta^W_Y+g_{YZ}\delta^W_X+
\Pi_{XZ}\Pi^W_Y-\Pi_{YZ}\Pi^W_X.  \label{ff}
\ee

{\tl Inverse fields:}
\be
\Pt_i^{X}=\left( \frac{\delta^a_b-\chi^a\chi_b}{1-\chi^2}\Et_i^b,
-\frac{ {\eps^a}_{bc}\Et_i^b\chi^c}{1-\chi^2} \right),
\qquad
\Qt_i^{X}=\left( \frac{ {\eps^a}_{bc}\Et_i^b\chi^c}{1-\chi^2},
\frac{\delta^a_b-\chi^a\chi_b}{1-\chi^2}\Et_i^b \right).
\label{invQ}
\ee

{\tl Projectors:}
\be
I_{(P)X}^Y =\left(
\begin{array}{cc}
\frac{\delta_a^b-\chi_a\chi^b}{1-\chi^2} &
\frac{ {\eps_a}^{bc}\chi_c}{1-\chi^2} \\
\frac{ {\eps_a}^{bc}\chi_c}{1-\chi^2} &
-\frac{\delta_a^b\chi^2-\chi_a\chi^b}{1-\chi^2}
\end{array} \right),
\qquad
 I_{(Q)X}^Y =\left(
\begin{array}{cc}
-\frac{\delta_a^b\chi^2-\chi_a\chi^b}{1-\chi^2} &
-\frac{ {\eps_a}^{bc}\chi_c}{1-\chi^2} \\
-\frac{ {\eps_a}^{bc}\chi_c}{1-\chi^2} &
\frac{\delta_a^b-\chi_a\chi^b}{1-\chi^2}
\end{array} \right).
\label{Qi-Qi}
\ee

{\tl Properties of the inverse fields and the projectors:}
\beq
&\Pt_i^X=-\Pi^X_Y\Qt_i^Y,
\qquad
I_{(P)}^{XY}=-\Pi^X_Z I_{(Q)}^{ZW} \Pi_W^Y,&  \\
& \tQ^i_X\Qt_j^X=\delta^i_j.  \quad
\tP^i_X\Pt_j^X=\delta^i_j,
\qquad
\tQ^i_X\Pt_j^X=\tP^i_X\Qt_j^X=0. &
\\
& I_{(P)Z}^Y I_{(P)X}^Z= I_{(P)X}^Y, \qquad
I_{(Q)Z}^Y I_{(Q)X}^Z= I_{(Q)X}^Y,
\qquad
I_{(P)X}^Y +I_{(Q)X}^Y=\delta_X^Y, & \nonumber
\eeq
The projector $I_{(P)}$ projects on $\tP$ and $\Pt$, whereas $I_{(Q)}$
projects on $\tQ$ and $\Qt$.
For example, one has $I_{(P)X}^Y \tP^i_Y=\tP^i_X$, $I_{(P)X}^Y \tQ^i_Y=0$
and other similar relations.
Another two useful identities are
\be
f^{WYZ} I_{(P)W}^X \tQ^i_Y \tQ^j_Z =0,
\qquad
f^{WYZ} I_{(Q)W}^X \tQ^i_Y \tQ^j_Z = f^{XYZ} \tQ^i_Y \tQ^j_Z.
\ee

{\tl Commutator of two shifted connection:}
\be
\{ \SA^X_i(x), \SA^Y_j(y)\}_D =
\hf \( \Pi^X_{X'} \CM_{ij}^{X'Y}- \CM_{ij}^{XY'}\Pi_{Y'}^Y\)  \delta(x,y),
\plabel{commAA}
\ee
where\footnote{There is a sign mistake in the first term in \Ref{tldM} in the printed
version of \cite{Alexandrov:2005ar}.}
\be
\CM_{ij}^{XY}(x,y) =
- \hf\( \CV_{ij}^{XY,l}(x)\p_l^{(x)}+\CV_{ji}^{YX,l}(y)\p_l^{(y)}\)+\CW_{ij}^{XY}
\plabel{tldM}
\ee
and
\beq
\CV_{ij}^{XY,l} & =&
f^{XP}_Q\left[\tQ^l_P\left(
(\Qt\Qt)_{ij}I_{(Q)}^{YQ}+\Qt_i^Y\Qt_j^Q-\Qt_j^Y\Qt_i^Q\right)
+\delta_i^l I_{(Q)}^{YQ}\Qt_j^P\right],
\plabel{VVV} \\
\CW_{ij}^{XY} & = &\hf\( \CL_{ij}^{XY}+\CL_{ji}^{YX}\)
+\frac{g_{SS'}}{2}\(I_{(P)}^{XT}\CV_{ij}^{SY,l} +
I_{(P)}^{YT}\CV_{ji}^{SX,l}\)\Qt_n^{S'}  \p_l \tQ^n_T ,
\label{WWW} \\
\CL_{ij}^{XY} &=& f^{PQ}_Z \left[ \Qt_j^{X}\Qt_n^Y\Qt_i^Z+
(\Qt\Qt)_{in}\Qt_j^{X}I_{(Q)}^{YZ}+\Qt_i^Y\Qt_n^{X}\Qt_j^Z
\right. \label{LLL} \\
&& \quad \qquad  - \left.
\Qt_i^Y\Qt_j^{X}\Qt_n^Z+(\Qt\Qt)_{ij}\Qt_n^{X}I_{(Q)}^{YZ}-
\Qt_j^Y\Qt_n^{X}\Qt_i^Z\right] \tQ^l_P\p_l\tQ^n_Q
\nonumber \\
&+& f^{Q}_{ZP} \left[ \Qt_n^Y\Qt_j^P+
(\Qt\Qt)_{jn}I_{(Q)}^{YP}-\Qt_j^Y\Qt_n^{P}\right]
I_{(Q)}^{ZX}\p_i\tQ^n_Q
+f^Z_{PQ}\Qt_j^{X}\Qt_i^QI_{(Q)}^{YP}\p_l\tQ_{Z}^l .
\nonumber
\eeq
Since this operator is implied to act on $\delta(x,y)$,
the argument of the last term in \Ref{tldM} is not important.
The antisymmetry of the bracket is ensured by the antisymmetry property of the matrix
\Ref{VVV}
\be
\CV_{ij}^{XY,l} = - \CV_{ji}^{YX,l} ,
\label{symV}
\ee
which can be checked by straightforward calculations.

%
\section{Consistency conditions}
%
\label{ap_con}

In this appendix we are going to check that the relation \Ref{negcons}
can be considered as a strong equality on the extended phase space spanned by
$(\tPp,\tPm,\SAp,\SAm)$. In other words, one should prove that the Dirac brackets
\Ref{comsh} remain true if one substitutes $\SAm$ by the r.h.s. of \Ref{negcons}.
This can easily be done for the chiral components of the triad. Indeed, one obtains
\beq
\{{\Xm^a}_c\, \SAp_i^c + 2{\Gamma^{\smash{(-)}}}_i^a, \tPp^j_b\}_D & = &
\delta_i^j{\Xm^a}_b=\{\Am_i^a, \tPp^j_b\}_D,
\\
\{{\Xm^a}_c\, \SAp_i^c + 2{\Gamma^{\smash{(-)}}}_i^a, \tPm^j_b\}_D & = &
\delta_i^j{\Xm^a}_c{\Xp^c}_b=\delta_i^j\delta^a_b=\{\Am_i^a, \tPm^j_b\}_D.
\eeq

For the two anti-selfdual connections one finds
\beq
&& \{{\Xm^a}_c\, \SAp_i^c + 2{\Gamma^{\smash{(-)}}}_i^a,
{\Xm^b}_d\, \SAp_j^d + 2{\Gamma^{\smash{(-)}}}_j^b\}_D
\nonumber \\
&& \qquad\qquad
=2\({\Xm^a}_c\{\SAp_i^c,{\Gamma^{\smash{(-)}}}_j^b\}_D
+\{{\Gamma^{\smash{(-)}}}_i^a,\SAp_j^d\}_D{\Xp_d}^b\).
\plabel{comAm}
\eeq
To reproduce the commutator $\{\SAm_i^a,\SAm_j^b\}_D$,
one should prove that this expression vanishes. This is natural to expect since
${\Gamma^{\smash{(-)}}}_i^a$ is a connection compatible with the three-dimensional metric.
Therefore, for $\chi=0$ the vanishing of \Ref{comAm} reduces to the well known statement
that the Barbero connection is commutative.
To do the calculations, it might be easier to work in the explicitly Lorentz-covariant
formulation where the statement we need to prove becomes
\be
\Rm^X_{Z}\(
\{\SA_i^Z,\Gamma_j^{W}\}_D
+\{\Gamma_i^{Z},\SA_j^W\}_D
\)\Rm^Y_{W}=0.
\label{comAmcov}
\ee
Using the explicit expression for $\Gamma_i^X$ \Ref{gam}, the commutation relation
\Ref{commsa} and various properties from appendix \ref{ap_def}, the relation
\Ref{comAmcov} can be checked by tedious and lengthy calculations.

Finally, it remains to prove that
\be
\{\SAp_i^a,{\Xm^b}_c\,\SAp_j^c + 2{\Gamma^{\smash{(-)}}}_j^b \}_D=
i\(\Rp\cdot \CM_{ij}\cdot\Rm\)^{ab} ,
\label{comApAm}
\ee
where $\CM_{ij}$ is given in \Ref{tldM}. Again, it is more convenient to work in
the covariant notations. Then the statement to be proved reads
\be
2 \Rp^X_{X'}\{\SA_i^{X'},\Gamma_j^{Y'} \}_D \Rm^Y_{Y'}= i\Rp^X_{X'}
\CM_{ij}^{X'Y'}\Rm^Y_{Y'}.
\label{comApAmcov}
\ee
Note that the connection $\Gamma_i^X$ is related to the quantity \Ref{VVV} as
\be
\Gamma_i^X=-\hf \CV_{ji}^{WX,l}\p_l\tQ^i_W.
\plabel{GVVV}
\ee
Substituting this relation into \Ref{comApAmcov} and using the antisymmetry
property \Ref{symV}, one immediately reproduces the first term in \Ref{tldM}
with derivatives. The remaining terms lead to the following statement
\beq
&&
-i \Rp^X_{X'}\{\SA_i^{X'},\CV_{jn}^{Y'W,l}\}_D \p_l\tQ^n_W \Rm^Y_{Y'}
\nonumber
\\
&& \qquad = \hf \Rp^X_{X'} \( \CL_{ij}^{X'Y'}+\CL_{ji}^{Y'X'}
-g_{WS}I_{(P)}^{Y'T}\CV_{ij}^{X'W,l}\Qt_n^S\p_l\tQ^n_T
+I_{(Q)W}^{X'}\p_l\CV_{ij}^{WY',l}\)\Rm^Y_{Y'}.
\plabel{comAApm}
\eeq
The Dirac bracket in the l.h.s. can be easily evaluated and one obtains
\beq
&&
-i \Rp^X_{X'} \{\SA_i^{X},\CV_{jn}^{YW,l}\}_D \p_l\tQ^n_W \Rm^Y_{Y'}
\nonumber \\
&& \qquad =
\Rp^X_{X'}\( \(\CV_{ji}^{Y'W,l}\Qt_n^{X'}+\CV_{in}^{Y'W,l}\Qt_j^{X'} \)\p_l\tQ^n_W
-\CV_{jn}^{Y'W,l}\Qt_l^{X'}\p_i\tQ^n_W\) \Rm^Y_{Y'}.
\plabel{comAV}
\eeq
Using the explicit expressions for $\CV_{ij}^{XY,l}$ and $\CL_{ij}^{XY}$ from
\Ref{VVV} and \Ref{LLL}, one can show that the r.h.s. of \Ref{comAApm} and
\Ref{comAV} indeed coincide.

%
\section{Triad representation}
%
\label{ap_repr}

We want to check that the operators \Ref{prep_self}, \Ref{prep_anti} give
a representation of the algebra \Ref{comsh} satisfying the condition \Ref{comconop}.
The latter fact is completely trivial as soon as the scalar product
is defined with the trivial measure $\CD\tP\CD\overline{\vphantom{\tilde P}\tP}$.
The commutation relations with the triad operators $\widehat{\vphantom{\tilde P} \tPp}$
and $\widehat{\vphantom{\tilde P} \tPm}$ also trivially give the necessary results.
The only non-trivial check must be done for the commutation relations
involving the connections.
For the two selfdual connections one finds
\beq
[\widehat{\vphantom{\tilde A} \SAp_i^a},\widehat{\vphantom{\tilde A} \SAp_j^b}]
&=&
i\(\frac{\delta{\Gamma^{\smash{(+)}}}_j^b}{\delta\tP^i_a}
+{\Xp^a}_c\,\frac{\delta{\Gamma^{\smash{(+)}}}_j^b}{\delta\overline{\vphantom{\tilde P}\tP^i_c}}
-\frac{\delta{\Gamma^{\smash{(+)}}}_i^a}{\delta\tP^j_b}
-{\Xp^b}_c\,\frac{\delta{\Gamma^{\smash{(+)}}}_i^a}{\delta\overline{\vphantom{\tilde P}\tP^j_c}}\)
\nonumber \\
& =& i\(\{\SAp_i^a, {\Gamma^{\smash{(+)}}}_j^b\}_D
+\{{\Gamma^{\smash{(+)}}}_i^a,\SAp_j^b, \}_D \).
\eeq
Then the property \Ref{propchg} and the constraint \Ref{negcons}
allow to rewrite this as
\be
[\widehat{\vphantom{\tilde A} \SAp_i^a},\widehat{\vphantom{\tilde A} \SAp_j^b}]
=\frac{1}{2i}{\Xp^a}_c\{\SAm_i^c, \SAm_j^d\}_D {\Xm_d}^b=0.
\ee
It is clear that the same result is valid for the two anti-selfdual connections.
In a similar way the commutator of the selfdual and the anti-selfdual connections
is found as
\beq
[\widehat{\vphantom{\tilde A} \SAp_i^a},\widehat{\vphantom{\tilde A} \SAm_j^b}]
&=&
i\(\frac{\delta{\Gamma^{\smash{(-)}}}_j^b}{\delta\tP^i_a}
+{\Xp^a}_c\,\frac{\delta{\Gamma^{\smash{(-)}}}_j^b}{\delta\overline{\vphantom{\tilde P}\tP^i_c}}
-\frac{\delta{\Gamma^{\smash{(+)}}}_i^a}{\delta\overline{\vphantom{\tilde P}\tP^j_b}}
-{\Xm^b}_c\,\frac{\delta{\Gamma^{\smash{(+)}}}_i^a}{\delta\tP^j_c}\)
\nonumber \\
& =& i\(\{\SAp_i^a, {\Gamma^{\smash{(-)}}}_j^b\}_D
+\{{\Gamma^{\smash{(+)}}}_i^a,\SAm_j^b, \}_D \)
 \\
& =& \frac{i}{2}\(\{\SAp_i^a, \SAm_j^b\}_D
-\{{\Xp^a}_c\SAm_i^c-\SAp_i^a, \SAm_j^b\}_D\)=i\{\SAp_i^a,\SAm_j^b, \}_D.
\nonumber
\eeq
This completes the proof that the operators \Ref{prep_self}, \Ref{prep_anti}
give a representation of the Dirac algebra.

\end{document}